\documentclass{article}

\usepackage{arxiv}

\usepackage[utf8]{inputenc} 
\usepackage[T1]{fontenc}    
\usepackage{hyperref}       
\usepackage{url}            
\usepackage{booktabs}       
\usepackage{amsfonts}       
\usepackage{amsmath}
\usepackage{amssymb}
\usepackage{bbold}          
\usepackage{nicefrac}       
\usepackage{microtype}      
\usepackage{graphicx}
\usepackage{natbib}
\usepackage{doi}
\usepackage[usenames,dvipsnames]{color}
\usepackage{xcolor}
\usepackage{dsfont}

\usepackage{listings}

\colorlet{lightorange}{orange!05}
\title{Evaluating the performance of {\sf npde} to evaluate joint models including longitudinal and TTE data: an application in metastatic hormono-resistant prostate cancer}
\date{May 4th, 2026}	

\author{ \hspace{1mm}Marc Cerou \\
Univ Rennes, Inserm, EHESP, \\
Irset (UMR S\_1085)\\
F-35000 Rennes \\
    \& Institut de Recherches Internationales Servier \\ 
    F-92150 Suresnes \\
    France \\
	\And
	\hspace{1mm}Jimmy Mullaert \\
Université Paris Cité \& \\
Université Sorbonne Paris Nord\\ 
Inserm, IAME, F-75018 Paris, France\\
\And
	\hspace{1mm}Marc Lavielle \\
    Inria Saclay \& Ecole Polytechnique \\ 
    F-91120 Palaiseau \\
    France \\
	\And
	\hspace{1mm}Sophie Peigné \\
    Institut de Recherches Internationales Servier \\ 
    F-92150 Suresnes, France \\
    \& Pharmatheus AB \\ 
    Uppsala, Sweden \\
	\And
	\hspace{1mm}Marylore Chenel \\
    Institut de Recherches Internationales Servier \\ 
    F-92150 Suresnes, France \\
    \& Pharmatheus AB \\ 
    Uppsala, Sweden \\
	\And
\href{https://orcid.org/0000-0002-9150-9886}{\includegraphics[scale=0.06]{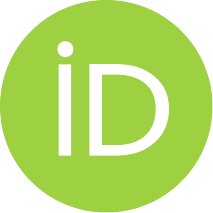}\hspace{1mm}Emmanuelle Comets}\thanks{Corresponding author} \\
Univ Rennes, Inserm, EHESP, \\
Irset (UMR S\_1085)\\
F-35000 Rennes \\
\& Université Paris Cité \& \\
Université Sorbonne Paris Nord\\ 
Inserm, IAME, F-75018 Paris, France\\
	\texttt{emmanuelle.comets@inserm.fr} \\
    }

\renewcommand{\headeright}{npde to evaluate joint models in NLMEM}
\renewcommand{\shorttitle}{{\sf arXiv} preprint}

\hypersetup{
pdftitle={npdeJointModels},
pdfauthor={Marc Cerou, Jimmy Mullaert, Marc Lavielle, Sophie Peigné, Marylore Chenel,Emmanuelle Comets},
pdfkeywords={Non-linear mixed effect models, Joint model, Model evaluation, npde, PSA, Prostate cancer},
}

\begin{document}
\newcommand{\un}{\mathds{1}}
\maketitle

\paragraph{Corresponding author:} Emmanuelle Comets \\
\begin{tabular}{p{3.2cm} l}
&INSERM - UMR1137 \\
&UFR de Médecine Site Bichat \\
&16 rue Henri Huchard \\
& 75018 Paris, France \\
&Tel: (33) 6 25 82 49 50\\
&\texttt{emmanuelle.comets@inserm.fr} \\
\end{tabular}




\newpage
\begin{abstract} 
{\bf Background and Objectives:}  Joint models are increasingly used in clinical trials. An important part of model building is to properly assess the descriptive and predictive ability of these models. Normalised prediction discrepancies (npd) and normalised prediction distribution errors (npde) have been developed to evaluate graphically and statistically non-linear mixed effect models for continuous responses. In this work, we propose to use a combined test to evaluate joint models, based on the p-values of the tests on longitudinal and time-to-event (TTE) data with a Bonferroni correction.

{\bf Methods:} Let V denote a dataset including both longitudinal and TTE observations. The null hypothesis H0 is that observations in V can be described by a model. Prediction discrepancies (pd) are defined as the quantile of the observation within its predictive distribution. In nonlinear mixed effect models (NLME), the predictive distribution is approximated by Monte-Carlo simulations. The pd for unobserved (censored) event times are imputed in a uniform distribution based on the model prediction of the probability of censoring, using a similar method as the one developed to handle data under the lower quantification limit (LOQ). Under H0, the pd follow a uniform U(0,1) and we normalise them to npd-TTE. In joint models, we compute separately the pd for TTE data and the prediction distribution error (pde) for the longitudinal data, which are obtained after decorrelating simulated and observed data, then normalised to npde. We then propose to use a combined test, combining the p-values of the tests on longitudinal data and on TTE data, adjusted with a Bonferroni correction.

We evaluated the performance of npd/npde through simulation studies based on a joint model developed by Desmée et al to characterise the relationship between the prostate specific antigen biomarker (PSA) and survival in 500 prostate cancer patients. We simulated event times and PSA trajectories from the joint model, for different sample sizes (50, 100, and 200), with a study duration of 1 year and 9 PSA samples. We evaluated the type I error and power of npd/npde to detect different types of model misspecifications, reflecting errors in the PSA model, in the hazard function or in the link function.

{\bf Results:} Plots of detrended npd-TTE were used to detect model misspecifications. In the simulation study, the npde-PSA were able to detect misspecifications in the PSA model, with a type I error close to 5\%. A misspecification on an influential parameter of the PSA model was captured by both npde-PSA and npd-TTE. This suggests that, if a test rejects the survival model, we have to look at whether the problem may not come from the longitudinal model.

For all types of misspecifications, the type I error of the combined test was found to be close to the expected 5\%. The power of the combined test to detect model misspecifications increased with the difference from the true model and as expected, with sample size. Graphically the power increase can be related to larger differences in the shape of the survival function or PSA evolution.

{\bf Conclusions:} npd can be readily extended for event data by imputing the pd for censored event under the model [1]. TThe difficulty with joint models is that events censor the longitudinal outcome so that the marker is not observed after the event occurs, which required an adjustment to compute the predictive distributions. Combining the tests on different outcomes using a Bonferroni correction is a general approach which led to good performance for the combined test for multiple response: the test showed an adequate type I error, and was quite sensitive to alternative models tested. 

\end{abstract}



\newpage
\section{Introduction}

\hskip 18pt Pharmacokinetic and pharmacodynamic models are used to extract the information in the data to support decision, for example the identification of a population at risk, or to propose an adapted dosage. In essence, all models make underlying assumptions which need to be evaluated and a critical step is to assess the adequacy between the model and the data~\cite{yano_evaluating_2001, comets_model_2010}. Nonlinear mixed effect models (NLMEM) have been widely used for analysing measurements collected in preclinical and clinical studies~\cite{nguyen_model_2017}. They allow to capture several level of variabilities, both intra- and between-subject and account for missing values and unbalanced data. Nguyen and al.~\cite{nguyen_model_2017} summarised different terminologies by separating "Model building" which is the process of developing a model on a given dataset, and "Model qualification" which is the assessment of the performance of a model to fulfill the analysis objectives. "Model evaluation" is required for both processes by governmental agencies~\cite{food_and_drug_administration_guidance_2003,agency_guideline_2007} and is an important part in model development~\cite{brendel_are_2007}. For this purpose and in the context of continuous data, several tools have been developed including the normalised prediction distribution errors (npde)~\cite{brendel_metrics_2006}, a decorrelated version of the normalised prediction discrepancies (pd)~\cite{mentre_prediction_2006}, and visual predictive checks (VPC)~\cite{holford_visual_2005, karlsson_tutorial_2008} which are considered as gold standards in the white paper written by the committee for best practices on evaluation of the International Society of Pharmacometrics~\cite{nguyen_model_2017}. 

In some clinical trials, the principal outcome can also be an event of interest, for example death, relapse or hospitalisation. Often, biomarkers are also recorded over time (e.g. tumor size or prostate specific antigen (PSA) in cancer trials). Joint models are defined as the simultaneous description of longitudinal and time-to-event (TTE) processes. They provide a promising statistical framework to estimate the association between those outcomes~\cite{henderson_joint_2000,rizopoulos_joint_2012, desmee_using_2017} and thus support clinical decisions and treatment choices. The emergence of estimation methods based on the likelihood maximisation with stochastic approximation of EM ({\sf SAEM}) or first-order conditional estimation with interaction ({\sf FOCEi}) and Laplace algorithms allow unbiased estimates of model parameters in joint models under appropriate missingness conditions~\cite{bjornsson_performance_2015,desmee_using_2017}. Since then, NLMEM have become increasingly more sophisticated, and joint models are increasingly used~\cite{sudell_joint_2016,tardivon_association_2019}. In a previous work, we extended the prediction discrepancies ($pd$) and their normalised version, called normalised predictive distribution ($npd$) to models describing time-to-event data, and showed their good performance to detect various types of model misspecifications in a simulation study considering only the TTE model~\cite{cerou_development_2018}. 

In the present paper, our objective is to propose $npd$ and $npde$ for joint models including both a longitudinal marker and time-to-event data, evaluate the performance of the statistical test based on these metrics and propose diagnostic graphs to visually diagnose model deficiencies. 
 
\newpage
\section{Models and methods}

\subsection{Statistical model}

\subsubsection{Model for longitudinal continuous outcome}

Let $Y_{ij}$ be the random variable associated with the $j$th longitudinal measurement of a biomarker for subject $i$ observed at time $t_{ij}$. We define $y_i$ as the vector of longitudinal measurements $\{y_{i1},y_{i2},\cdots,y_{in_i}\}$ observed for subject $i$ with $n_i$ the number of measurements for that subject. 

We assume that the observations can be described by the following model:
\begin{equation}
\begin{split}
    y_{ij} &= f(t_{ij},\psi_i) + g(t_{ij},\psi_i, \sigma)\epsilon_{ij} \\
    \psi_i &= h(\mu, z_i,\eta_i) \\
\end{split}
\end{equation}
where the structural model $f$ is a known (possibly nonlinear) function of design variables $t_{ij}$ and individual parameters $\psi_i$, $g$ represents the standard deviation of residual errors multiplying $\epsilon_{ij} \sim \mathcal{N}(0,1)$. The parameters are expressed as a function $h$ of a linear combination of fixed effects $\mu$, the typical value in the population, subject-specific covariates $z_i$ and individual random effects $\eta_i$. The function $h$ is typically a normal, log-normal, or logit-normal distribution, and $\eta_i$ are assumed to follow a normal distribution ($\eta_i \sim \mathcal{N}(0,\Omega)$ with $\Omega$ the variance-covariance matrix).

\subsubsection{Time-to-event model}

We denote $T_i$ the time to event observed in subject i. To avoid confusion with the design variable $t_{ij}$ representing the sampling times for the continuous measurements, we will denote $T$ the random variable associated to an event/censoring time and keep a capital letter (T$_i$) for the realisation of this random variable in subject $i$. We assume that $T_i$ lies within the interval $T_{L_i}$ and $T_{R_i}$. This definition allows all types of censoring processes as well as not censored event. Indeed, if the event is not censored, $T_{L_i} = T_{R_i} = T_i$. If the event is right censored, by the end of the study for example, $T_{R_i}=\infty$ and if the event is left censored, $T_{L_i}=0$. In any case, the instantaneous risk of event is described through the parametric hazard function:

\begin{equation}
    h(t) = \lim\limits_{dt \rightarrow 0^+} \frac{Pr(t\le T<t+dt|T\ge t)}{dt}
\end{equation}

The probability to be event-free $S$ at time $t$ is obtained by integrating the hazard until time $t$:

\begin{equation}
    S(t) = \exp\left(-\int_0^t h(x)dx\right)
\end{equation}

\subsubsection{Joint model with longitudinal and TTE data}

In clinical trials, multiple outcomes of interest can be recorded and an important goal is to assess the association between those outcomes. 
In this paper, we assume conditional independence between outcomes (longitudinal measurement and TTE data) through a shared-random effect model~\cite{wulfsohn_joint_1997}. In this model, the hazard function is traditionally decomposed into a parametric baseline hazard $h_0(t)$ and a second component characterising the association between the risk and a longitudinal outcome, as well as an association with covariates, as in:
\begin{equation}
    h(t) = h_0(t) \: e^{\beta X(t, \psi_i, z_i) + \alpha z_i} \label{eq:hazardJointTTE}
\end{equation}
where $X(t, \psi_i, z_i)$ is an association function (which can be equal to $f$, see possible choices in the simulation study), $\beta$ measures the strength of the association with the longitudinal marker, and $\alpha$ measures covariate effects. We define the hyperparameters $\theta$ as composed by the fixed effects, covariate and association effects, error model parameters and the variance-covariance matrix: $\theta = \{\mu, \sigma, \Omega\, \beta, \alpha \}$.

The conditional independence assumption means that the joint probability of an observation $Y_{ij}$ at time $t_{ij}$ and the event $T_i$ conditional to the individual parameters is the product of the two conditional probabilities $p_i(Y_{ij},T_i|\eta_i, \theta) = p_i(Y_{ij}|\eta_i;\theta) \; p_i(T_i|\eta_i;\theta)$. Because the individual random effects $\eta_i$ are unknown, it follows that the joint probability by integrating over the distribution $\mathbb{R}^d$ of the individual parameters as: 
\begin{equation}
    p_i(Y_{ij},T_i|\theta)=\int p_i(Y_{ij},T_i|\eta_i;\theta)p_i(\eta_i;\theta)d\eta_i = \int p_i(Y_{ij}|\eta_i;\theta) \; p_i(T_i|\eta_i;\theta) \; p_i(\eta_i;\theta)d\eta_i
\end{equation}

\subsection{Model evaluation with {\sf npde}}

\subsubsection{{\sf npd} and {\sf npde} for single-response continuous data}

For continuous data, Mentré and Escolano~\cite{mentre_prediction_2006} proposed to compute for each observation $y_{ij}$ the associated prediction discrepancy, denoted $pd_{ij}$, which corresponds to the value of the cumulative (marginal) distribution function (cdf) of the predictive distribution for that observation:

\begin{equation}
    pd_{ij} = \mathbb{P}(Y_{ij}<y_{ij}| \theta) = \int
    \mathbb{P}(Y_{ij}<y_{ij}|\eta, \theta) p(\eta|\theta) d\eta \label{eq:pdcontinu}
\end{equation}

The $pd$ is indexed by $ij$ to denote the dependency on time $t_{ij}$, subject-specific design variables and covariates. Because the individual parameters are unknown, and expression~(\ref{eq:pdcontinu}) does not have a closed form, the quantity $pd_{ij}$ is approximated by Monte-Carlo simulation. More precisely, if we denote $\eta_i^{sim(k)}, k=1,\ldots,K$, a random vector drawn from the distribution $p(\eta_i|\theta)$ and $y_i^{sim(k)}=\{y_{i1}^{sim(k)}, \ldots, y_{i\; n_i}^{sim(k)} \}$, the resulting vectors of simulated observations from the model, we can estimate $pd_{ij}$ by:
$$
\widehat{pd}_{ij} = \dfrac{1}{K} \sum_{k=1}^K \mathbb{1}_{y_{ij}^{sim(k)}<y_{ij}},
$$
where $\mathbb{1}_{C}=1$ if condition $C$ is satisfied and $0$ otherwise.

If the observation $y_{ij}$ is drawn from the correct marginal distribution, the inverse cdf theorem implies that the corresponding $\widehat{pd}_{ij}$ follows a uniform distribution $\mathcal{U}(0,1)$ when $K \to \infty$. To ease the interpretation by comparison with classical residuals, $pd$ are usually normalised to $npd$ which then follow a normal distribution $\mathcal{N}(0,1)$. 

However, because observations within individuals are correlated and therefore not independent, the $npd$ are correlated within each individual. In the context of model evaluation, where we are interested in testing the empirical distribution of $npd_{ij}$ against a normal distribution, this leads to a type I error inflation as classical tests of normality assume independence between the $npd$~\cite{mentre_prediction_2006}. In other words, the model may be rejected due to this correlation and not because the model is wrong. 
Brendel et al.~\cite{brendel_metrics_2006} proposed to decorrelate the observations and the marginal predictive distribution and compute the quantile of a decorrelated observation in its decorrelated marginal predictive distribution, called prediction distribution error ($pde$). Normalised $pde$ are called normalised prediction distribution errors ($npde$). Under $H_0$, $pde$ follow a uniform distribution $\mathcal{U}(0,1)$ and $npde$ a normal distribution $\mathcal{N}(0,1)$. 

\subsubsection{{\sf npd} for TTE data in the context of a joint model}

\hskip 18pt For time-to-event data, we generalised in a previous work~\cite{cerou_development_2018} the approach proposed by Nguyen et al.~\citep{nguyen_extension_2012} for continuous data below the limit of quantification (BLQ) which can be viewed as left censored) to interval censored events in general. We define the cumulative marginal predictive distribution $F$ of a time variable $t$ as:

\begin{equation}
    F_i(T) = \int_0^T \int p_i(t|\eta;\theta)p_i(\eta|\theta) \; d\eta dt
    = \int_0^T \int h_i(t|\eta;\theta) S_i(t|\eta;\theta) p_i(\eta|\theta) \; d\eta dt \label{eq:pdTTE}
\end{equation}

When the event is observed at time $T_i$, the $pd$ corresponds to the same definition as in the continuous case as $pd_i = F_i(T_i)$. When the event is censored, we have $T_{L_i}=T_{cens}$ and $T_{R_i}= \infty$ and we proposed to sample the $pd$ in a uniform distribution for an event which lies within the interval $[T_{L_i},T_{R_i}]$: $pd_i \sim \mathcal{U}\left(F_i\left(T_{L_i}\right), 1 \right)$ as $F_i\left(T_{R_i}\right)=1$. Note that this extends to interval censored events in general where we sample  $pd$ in a uniform distribution for an event which lies within the interval $[T_{L_i},T_{R_i}]$ as $pd_i \sim \mathcal{U}\left(F_i\left(T_{L_i}\right), F_i\left(T_{R_i}\right) \right)$. 

We showed the good performance in term of type I error and power of the tests evaluating the distribution of the $npd$ in the context of joint model with only TTE data and in several cases of model misspecification in~\cite{cerou_development_2018}. In that study we assumed a terminal event occurring only once, so there was no inflation due to repeated measurements.

\subsubsection{{\sf npd/npde} for continuous data in the context of a joint model}

\hskip 18pt For joint models of a continuous observation with a time-to-event process, the marginal predictive distribution of the continuous observation must take into account the probability that an observation exists or not~\cite{rizopoulos_joint_2012,tsiatis_joint_2004}. To formalise this, assume that $n_{tot,i}$ observations have been planned for subject $i$ (with $n_i$ the number of samples actually collected), and denote $y_{tot,i} = \{ y_{i1}, \ldots, y_{i n_{tot,i}} \}$ be the vector of complete observations planned at times $t_{tot,i} = \{ t_{i1}, \ldots, t_{i n_{tot,i}} \}$. $t_{i, n_i} \leq T_i$ is then the time of the last observation actually collected ($n_i \leq n_{tot,i}$, with an inequality when censoring occurs before the end of the sampling planned). Expression~(\ref{eq:pdcontinu}) is applicable to the complete set of observations $y_{tot,i}$, but the cumulative distribution function for responses actually observed is conditional on time being before the event or censoring time.

The conditional independence assumption lets us write the prediction discrepancy with respect to the correct marginal predictive distribution as:
\begin{equation}
\begin{split}
    pd_{ij} 
    &= \mathbb{P}(Y_{ij}<y_{ij}|T_{i}>t_{ij},\theta) \\
    &= \dfrac{\mathbb{P}(Y_{ij}<y_{ij},T_{i}>t_{ij}|\theta)}{\mathbb{P}(T_{ij}>t_{ij},\theta)} \\ \label{eq:conditionaldensity}
    &= \dfrac{\int p(Y<y_{ij}|\eta,\theta) \; p(T>t_{ij}|\eta,\theta) \; p(\eta|\theta) \; d\eta}
    {\int p(T>t_{ij}|\eta,\theta) \; p(\eta|\theta) \; d\eta}\\
\end{split} 
\end{equation}

In practice, $K$ replicates for each vector of observations are simulated from the tested model. For each $k$ replicate ($k=\{1,\cdots,K\}$), $\eta_i^k$ are sampled in $\mathbb{R}^d$ and we define the predicted TTE value $T_i^{sim(k)}$, and the vector of predicted longitudinal observations $y_{ij}^{sim(k)}$ observed at time $t_{ij}$. 

An unbiased estimator of the $pd_{ij}$ for each observation is computed as a ratio of Monte-Carlo estimation:  

\begin{equation}
    pd_{ij} = \frac{\sum_{k=1}^K \mathbb{1}_{y_{ij}^{sim(k)} < y_{ij}}\times \mathbb{1}_{T_{i}^{sim(k)}>t_{ij}}}
    {\sum_{k=1}^K \mathbb{1}_{T_{i}^{sim(k)}>t_{ij}} }
\end{equation}

In order to decorrelate the $pd$ for continuous data, Brendel et al.~\cite{brendel_metrics_2006} proposed to decorrelate the observations and predictions before the computation of the $pd$ by evaluating the mean $E(Y_i)$ and the variance-covariance matrix $Var(Y_i)$ between the observation time $t_{ij}$ over the $K$ simulations under the tested model. We use the same method here to decorrelate the $pd$ associated with the continuous observations. The tested model allows the estimation of the mean regardless of the simulated event time:
\begin{equation}
    \bar{Y}_i = \frac{1}{K} \sum_{k=1}^K y_i^{sim(k)}
\end{equation}

and the variance-covariance matrix:
\begin{equation}
    V_i = \frac{1}{K-1} \sum_{k=1}^K \left(y_i^{sim(k)} - \bar{Y}_i \right)\left(y_i^{sim(k)} - \bar{Y}_i \right)'
\end{equation}

Both observed and simulated data are decorrelated as follows:

\begin{equation}
    y_i^* = V_i^{-\frac{1}{2}} \left(y_i - E(Y_i)\right)
\end{equation}

\begin{equation}
    y_i^{sim(k)*} = V_i^{-\frac{1}{2}} \left(y_i^{sim(k)} - E(Y_i)\right)
\end{equation}

The decorrelated version of the $pd$, called $pde$, are computed as :

\begin{equation}
    pd_{ij} = \frac{1}{\sum_{k=1}^K \un_{t_{ij} < T_i^{sim(k)}} } \sum_{k=1}^K \un_{y_{ij}^{sim(k)*} < y_{ij}^*}\times \un_{t_{ij} < T_{i}^{sim(k)}}
\end{equation}

$npd$ and $npde$ are obtained by a normalisation of the $pd$ and $pde$ using the inverse cumulative normal distribution function. 



\subsubsection{Tests for npde in joint models}

\hskip 18pt Let $npd_L/npde_L$ denote the $npd$/$npde$ for the longitudinal outcome. Similarly let $pd_{TTE}$/$npd_{TTE}$ denote the metrics for the TTE outcome. Under the null hypothesis $H_0$, $npde_L$ and $npd_{TTE}$ follow a normal $\mathcal{N}(0,1)$ distribution. To test the null hypothesis, \citep{brendel_metrics_2006} used a global test combining a Wilcoxon signed rank for the mean, a Fisher test for the variance, and a Shapiro-Wilk test for normality corrected using a Bonferroni correction. In the present work we apply both the global test and an adjusted Kolmogorov-Smirnov test. For the global test, we apply the three tests again separately to the $npde_L$ and $npd_{TTE}$, resulting in 6 p-values (3 for the longitudinal part and 3 for the TTE part), which are corrected using Bonferroni correction, so that the model is rejected if the lowest p-value is lower than 0.05/6. For the Kolmogorov-Smirnov test, we compare separately the $npde_L$ and $npd_{TTE}$ to $\mathcal{N}(0,1)$, and apply a Bonferroni correction, rejecting the model if the lowest p-value is lower than 0.05/2.

\subsubsection{Graphs for joint model diagnostics with npde}

\hskip 18pt The graphical evaluation of the model for continuous outcomes recommended in~\cite{nguyen_model_2017} are based on the evaluation of 
scatterplots of $npd_L$ versus time, predictions, or covariates. In these graphs, the median of the $npd$ and selected (usually 5-95$^{\rm th}$) percentiles at successive values on the X-axis (possibly binned over intervals) are compared to the 90 or 95\% prediction interval of the associated theoretical percentiles. The distribution of the $npd$ can also be compared to the theoretical normal distribution through quantile-quantile plots and histograms. These different graphs can be obtained for example in Monolix~\cite{lavielle_book} or in the {\sf npde} library~\cite{comets_computing_2008}.


For TTE data, because only one outcome (event or censoring) is observed for each individual, scatterplots are less informative, since $npd_{TTE}$ are population residuals and should be expressed as a function of the predicted time-to-event in the population (not the individual event time), which only influenced by covariates in the absence of variability. Therefore, Cerou et al. recommended quantile-quantile plots and histograms of $npd_{TTE}$ to assess this component of the model~\cite{cerou_development_2018}. Another diagnostic which attempts to detect a model misspecification in the time-scale of the study is a wormplot~\cite{vanbuuren_wormplot_2001}, computing de-trended residuals. We first determine for each $pd_{TTE,i}$ a theoretical quantile with its 90\% prediction interval computed based on the Clopper-Pearson method~\cite{clopper_use_1934}, also called the exact method as it is based on the correct distribution rather than an approximation. A de-trended value can be obtained by subtracting the $pd_{TTE,i}$, the lower and upper limit of the prediction interval to the theoretical value. In this way, the de-trended value of $pd_{TTE}$ can be expressed over time with a prediction interval. The de-trended prediction interval indicates only if the de-trended $pd_{TTE}$ is different from 0. A deviation from this interval however can indicates an over/under-prediction of the model and can be directly interpreted as a difference in the probability of event.




\subsection{Evaluation of the performance of the test}

We performed a simulation study to evaluate the performance of the test based on joint $npde_L$ and $npd_{TTE}$ to detect model misspecifications related to different components of the joint model. The context was the same as in the previous work~\cite{cerou_development_2018} but here considering both the longitudinal and TTE data and not only the TTE data. We first describe briefly the motivating example and follow up with the simulation study design and evaluation.

\subsubsection{Motivating example}

Patients with metastatic hormono-resistant prostate cancer are at high risk of death. Clinical trials evaluating treatment often monitor biomarkers which are linked to survival in order to assess treatment efficacy at an earlier date compared to survival time. Desmée et al.~\cite{desmee_nonlinear_2015} developed, based on the VENICE phase III clinical trial, a semi-mechanistic model to describe the evolution of prostate specific antigen (PSA) in patients given chemotherapy, and modelled its association with the TTE parametric model. In the simplified version of the model used for a later simulation study~\cite{desmee_using_2017}, the evolution of PSA was described by a bi-exponential model: at treatment initiation, PSA start at a baseline value, $PSA_0$, and then decreases at a rate which depends on the treatment efficacy parameter $\epsilon$. After a certain time point, $T_{esc}$, cancer cells escape from the treatment and start to grow at a rate $r$, producing PSA. The equation for PSA includes 4 parameters:
\begin{equation}
PSA(t, \theta) = 
\left\{
\begin{matrix}
\frac{\delta PSA_0}{r(1-\epsilon) - k_{out} + \delta} \; e^{(r(1-\epsilon) - k_{out}) \; t}
+ \left( PSA_0 - \frac{\delta PSA_0}{r(1-\epsilon) - k_{out} + \delta} \right) \; e^{-\delta \;t}
& \text{if } t \leq T_{esc} \\
\frac{\delta PSA_0}{r(1-\epsilon) - k_{out} + \delta} \; e^{(r - k_{out}) \; t - r \epsilon T_{esc}} + \left( PSA(T_{esc}) - \frac{\delta PSA_0 \; e^{(r(1-\epsilon) - k_{out}}) T_{esc}}{r - k_{out} + \delta} \right) \; e^{-\delta \; (t-T_{esc})} 
& \text{otherwise}
\end{matrix}
\right.
\end{equation}
The constants $k_{out}$ and $\delta$, reflecting respectively  the rate of prostate cell elimination and the rate of PSA elimination, were fixed to 0.046 day$^{-1}$ and 0.23 day$^{-1}$ as in the original paper~\cite{desmee_nonlinear_2015}.

The hazard for the survival depends on a Weibull function for the baseline hazard, and we assume that the hazard depends on the current predicted value of PSA ($X(\psi_i,t)=PSA(\psi_i,t)$ in equation~\ref{eq:hazardJointTTE}) through the following function:
\begin{equation}
    h_i(t,\psi_i) = \frac{k}{\lambda}\left(\frac{t}{\lambda}\right)^{k-1} \times \exp\left(\beta\cdot PSA(\psi_i,t)\right)
\end{equation}
where $k$ is the shape of the Weibull baseline hazard model and $\lambda$ the scale. In this parametrisation $\lambda$ corresponds to the time when the probability to be event-free equals to 0.37. 

The parameters used in the simulation study for the base model were the same as in~\citep{cerou_development_2018} and are given in Table~\ref{psa_paramter}, along with their distribution for the parameters of the PSA submodel. No interindividual variability (IIV) was assumed for the TTE model as it would not be identifiable from only one event per subject.

\subsubsection{Simulation study and settings}

We evaluated the performance of the adjusted Kolmogorov-Smirnov and global test described in the test section through three simulations studies. The context of these simulations can be viewed as an external evaluation where we want to evaluate the descriptive ability of a model (defined as the structural and statistical model, as well as the parameters) on new data (for example a model built on phase II clinical trials and evaluated on phase III clinical trials). $K=2000$ replicates were used in the Monte-Carlo simulation used to approximate the marginal predictive distribution. 

First, we tested the ability of the tests to detect a misspecification on the baseline hazard model and more specifically on the shape parameter $k$. Data were generated from different true values: $k=1$, $k=1.5$. In each case, the marginal predictive distribution of each outcome was computed from models with $k=0.8$, $k=1$, $k=1.2$, $k=1.5$, $k=2$. The type I error is obtained when true and tested model match, while in the other cases we evaluate the power to detect each model misspecification.

Second, we simulated two types of misspecifications in the model for the longitudinal outcome: one in the treatment efficacy parameter $\epsilon$ in the PSA model and one in the standard deviation of the inter-individual variability of this parameter, $\omega_{\epsilon}$. The true value of $\epsilon$ used to generate data was $\epsilon=0.3$ (mild efficacy) and we tested 4 different values: $\epsilon=0.15$ (low efficacy), $\epsilon=0.3$ (mild efficacy), $\epsilon=0.45$ (moderate efficacy) and $\epsilon=0.8$ (high efficacy). The true value of $\omega_{\epsilon}$ was 1.5 and we tested 3 different values: $\omega_{\epsilon}=0.6$, $\omega_{\epsilon}=1$, $\omega_{\epsilon}=1.5$. 
Third, we simulated a misspecification in the association function $X$ from equation~\ref{eq:hazardJointTTE}. We generated data through different true models $M_V$ assuming the hazard is impacted by the current value of PSA($X(t,\psi_i) = PSA(t,\psi_i)$) or the slope of the logarithm of PSA ($X(t,\psi_i) = \frac{dPSA(t,\psi_i)}{dt}$). We then evaluated 6 possible models (shown in Table~\ref{tab:models}), keeping the same parameter values as in the base model.


\subsubsection{Performance evaluation}

In all scenarios, data was generated with three sample size ($N = \{50, 100, 200\}$) and a maximum number of measurements of 9 for PSA, equally spaced on time scale from treatment initiation ($t=0$) to one year ($t=365$ days). To compute the type I error and power of each test, data for each scenario were generated 200 times. The type I error is expected to be 0.05 and to remain within the interval [0.024 - 0.09] (prediction interval for the exact Binomial test). 



\subsection{Implementation}
We used the statistical software R~\cite{r_core_team_r_2018}, version 3.5.1, to compute the $pd$ for TTE data. For $npd_L/npde_L$ adapted to continuous data, we extended code from the R package {\sf npde}~\cite{comets_computing_2008} to compute the $npd_{TTE}$. A self-contained example with simulated data was uploaded to a Zenodo depot (\url{10.5281/zenodo.10831899}) and is also available on the development github for npde (\url{https://github.com/ecomets/npde30/tree/main/cerouJoint}). For the simulation of continuous and TTE data, we used the function {\sf simulx} from the R package {\sf mlxR} providing an interface to Monolix~\cite{mlxR, lavielle_book}. 

To highlight the difference in event probability between models, we also computed the Kaplan-Meier VPC (KMVPC). In this representation, the summary measure is the Kaplan-Meier estimates of the event probability. Kaplan-Meier estimates is computed in the data and in each of the replicates of the Monte-Carlo simulations. At each binned time, the $5^{\rm th}$, $50^{\rm th}$, and $95^{\rm th}$ percentiles of the Kaplan-Meier estimation of event probability are computed. The overall model adequacy is evaluated by comparing the Kaplan-Meier event probability in the data to the prediction interval.  

\newpage
\section{Results}

\subsection{Using graphs and tests for model diagnostics with {\sf npde}}

\begin{figure}[h!]
    \centering
    \includegraphics[width=1\textwidth]{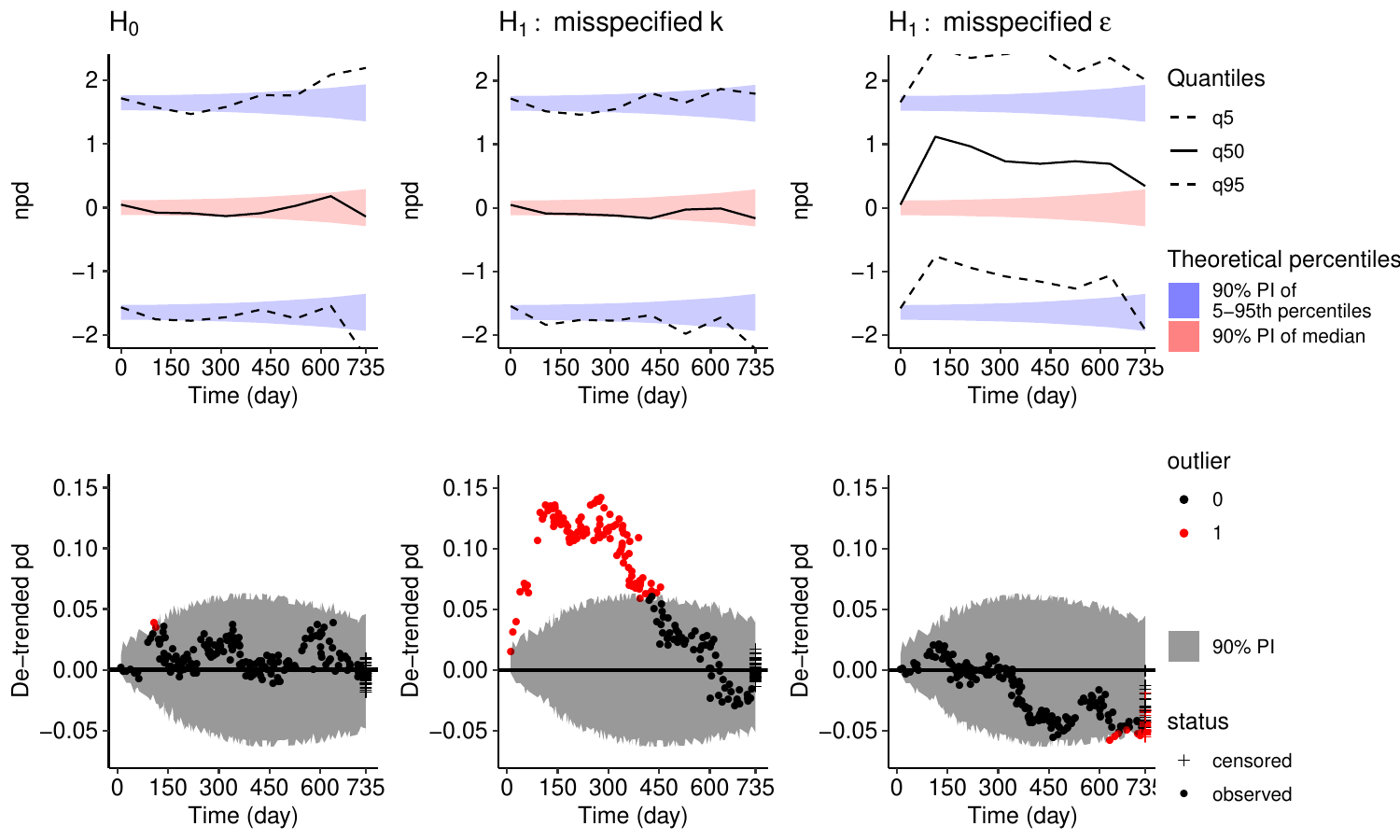}
    \caption{$npd$ for longitudinal data and de-trended $pd_{TTE}$ for TTE data over time under $H_0$ (left), with a misspecified baseline hazard function (middle: constant model instead of Weibull) and a misspecified treatment effect (right: $\epsilon=0.8$ instead of $\epsilon=0.3$). Top: the prediction bands correspond to the 90\% prediction interval of the theoretical quantiles. Bottom: The prediction band corresponds to the 90\% de-trended prediction interval. Dots within this band are not different from zero while dots outside are the signal of an over/under prediction.}
    \label{fig:npde_npd_H0_H1_k}
\end{figure}

Fig.~\ref{fig:npde_npd_H0_H1_k} shows example of graphs evaluating the longitudinal and TTE process under $H_0$ (left column) and under two alternative $H_1$ hypotheses (middle and right columns). On the left, the tested model is the base model with a Weibull hazard function ($H_0$) which was also used to generate the data; in the two other columns, the same data was used but the tested models were respectively a constant baseline hazard model ($H_1: k=1$) instead of the Weibull hazard (middle plots) and a model with a larger treatment effect ($H_1:  \epsilon=0.8$), from the second simulation scenario (right plots). The graphical evaluation consists in representing the $npd_L$ for the longitudinal PSA biomarker measurements (uppper row), and the de-trended $pd$ over time for the event (lower row). 

Under $H_0$, both for the longitudinal and TTE graphs over time, $npd_L$ quantiles are within the 90\% prediction interval of their associated theoretical quantiles. De-trended $pd_{TTE}$ are all within the 90\% prediction interval which indicates that they are not different from zero and thus the model characterise well the data. In joint models, the number of longitudinal observations decrease over time due to the time-to-event process. Thus, the theoretical percentiles and their prediction interval take into account the number of observations in each binned time. By construction, the size of the prediction intervals of the 5-95$^{\rm th}$ theoretical percentiles increases as the sample size decreases, hence the widening of the prediction intervals in the upper row as time increases in Fig.~\ref{fig:npde_npd_H0_H1_k}.

Under $H_1$ where the tested model is a constant hazard baseline model ($k=1$), quantiles of $npd_L$ are within the 90\% prediction interval of their associated theoretical quantiles so this submodel is not rejected. However, the model overpredicts the risk of event in the beginning (de-trended $pd_{TTE}$ quantiles over 0) and then the de-trended $pd_{TTE}$ tends to 0. This is expected as the true value of $k$ is 1.5 (Weibull model) so in this configuration, a constant risk ($k=1$) induces an increasing risk of events occurring at the beginning of the treatment period.

Under $H_1$, the misspecification is carried by the treatment efficacy parameter of the PSA model and impacts the model for the longitudinal marker. This is reflected in the top right plot, where the median and 90\% percentiles of $npd_L$ are above their theoretical prediction interval for all time except at baseline. This shows that the model underpredicts the PSA and there is a misspecification in the mean structure of the model. This is expected as the treatment efficacy in the tested model ($\epsilon=0.8$) is considerably higher than in the data ($\epsilon=0.3$). Despite the strong misspecification in the longitudinal submodel, all the de-trended $pd_{TTE}$ are within their de-trended prediction interval, so we can consider that the TTE-submodel describes well the event data.


\subsection{Performances of the test}

For the sake of simplicity, we present the results only for the adjusted global test as the adjusted Kolmogorov-Smirnov test presented similar performances in almost every case, excepted for one model misspecification (between subject variability in PSA submodel parameters in the second simulation). The full results of the simulation study are available as supplementary material. 

\subsubsection{Misspecification on TTE model parameters}

\begin{figure}[h!]
    \centering
    \includegraphics[width=0.9\textwidth]{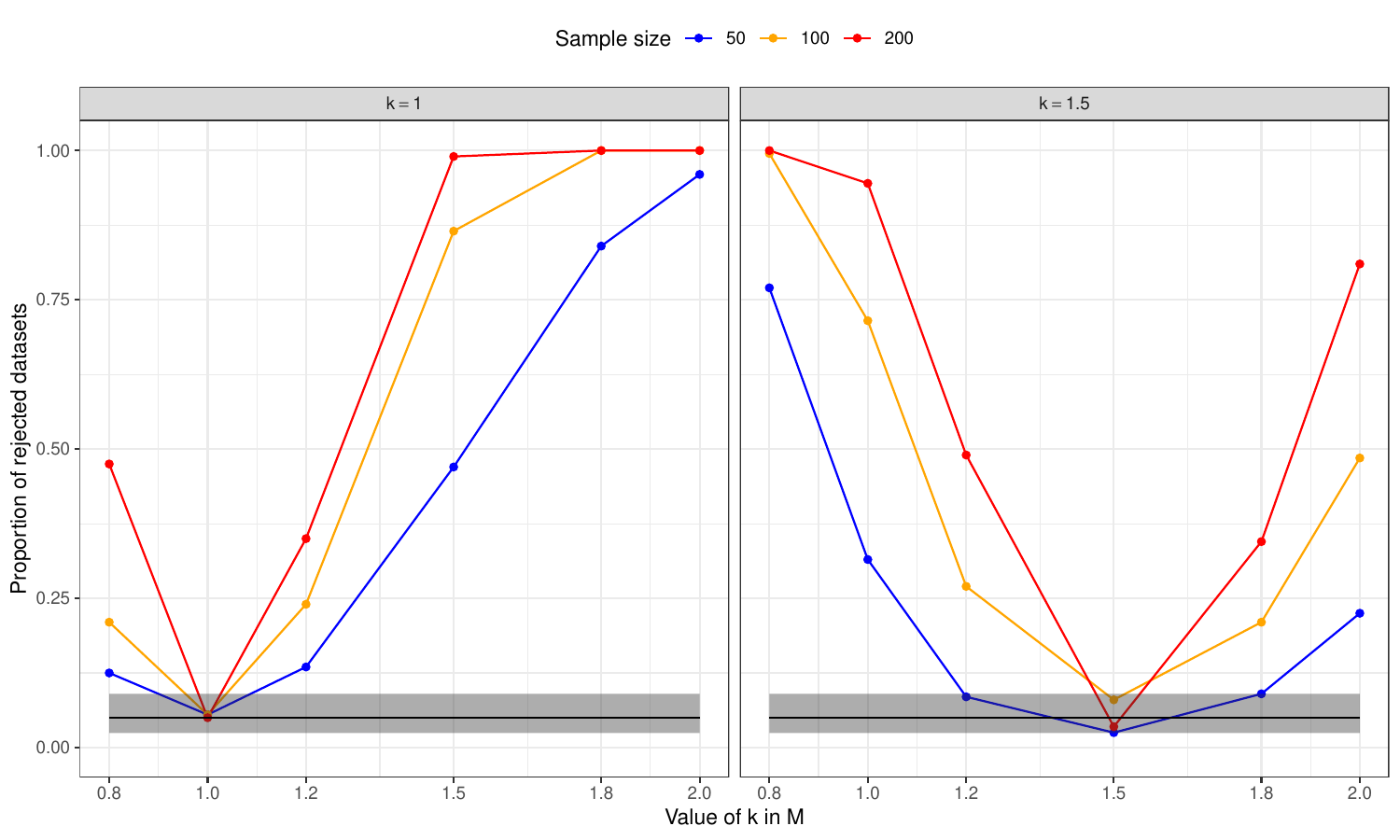}
    \caption{Performance in case of misspecifications on TTE submodel parameters. Data are generated under a constant baseline hazard model (k=1) on the left and a Weibull model (k=1.5) on the right. For each scenario, three sample size are tested (N=50 (blue), N=100 (orange), N=200 (red)).}
    \label{fig:power_k}
\end{figure}

\hskip 18pt Fig. \ref{fig:power_k} shows the type I error and power under different scenarios where the misspecification is on the shape parameter $k$ of the Weibull baseline hazard model. On the left, data are generated under a constant baseline hazard (exponential survival model, $k=1$). On the right, data are generated with an increasing baseline hazard over time ($k=1.5$). In both cases, the type I error is controlled within the prediction interval of the 5\%. The power increases as the sample size and the difference from the true value increase which is in accordance with what it is expected.

\subsubsection{Misspecification on PSA submodel parameters}

\begin{figure}[h!]
    \centering
    \includegraphics[width=0.9\textwidth]{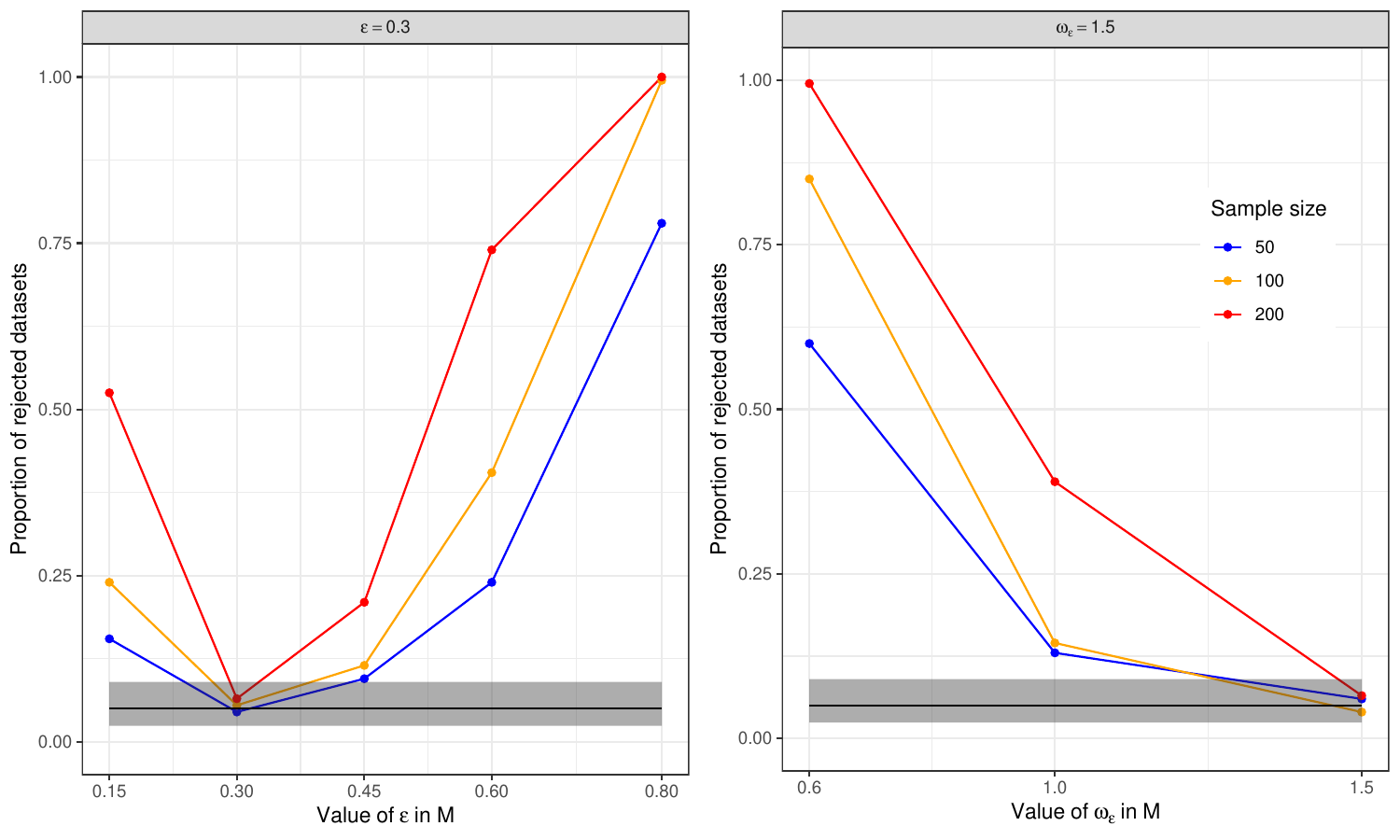}
    \caption{Performance in case of misspecifications on PSA submodel parameters. Left: performance of the global test over increasing values of treatment effect ($\epsilon$). Right: performance of the global test over increased values of the between subject variability on this parameter ($\omega_{\epsilon}$). For each scenario, three sample size are tested (N=50 (blue), N=100 (orange), N=200 (red)).}
    \label{fig:power_epsilon}
\end{figure}

\hskip 18pt Fig. \ref{fig:power_epsilon} shows the performance in case of a misspecification on the treatment efficacy parameter $\epsilon$ (left) or in its variability $\omega_{\epsilon}$ (right). In the first scenario, a mild treatment efficacy (right) was used to generate the data and 5 values of $\epsilon$ were tested, including the true value. Type I error is controlled and as before the power increases as the difference from the true value and the sample size increase. In the second scenario, $\omega_{\epsilon}=1.5$ was used to generate the data and 3 increasing magnitudes of variability were tested. The type I error is controlled within the prediction interval of 5\%. 


The scenarios where the simulated data had a different variability for $\epsilon$ or $T_{esc}$ compared to the tested model were the only cases where the global adjusted and adjusted Kolmogorov-Smirnov had very different powers, with the power of the adjusted Kolmogorov-Smirnov much lower than the global test. Fig. S6 and S8 in supplementary material shows both tests, and illustrates that the adjusted Kolmogorov-Smirnov has almost no power to detect this type of model misspecification even with the largest number of subjects where the power of the global test is 100\%. This may be due to a lack of sensitivity in the Kolmogorov-Smirnov test to the variance of the distribution, whereas in the global test we use a Fisher test for the variances which is more specific.

\subsubsection{Misspecified association between PSA and TTE submodels}

\begin{figure}[h!]
    \centering
    \includegraphics[width=0.9\textwidth]{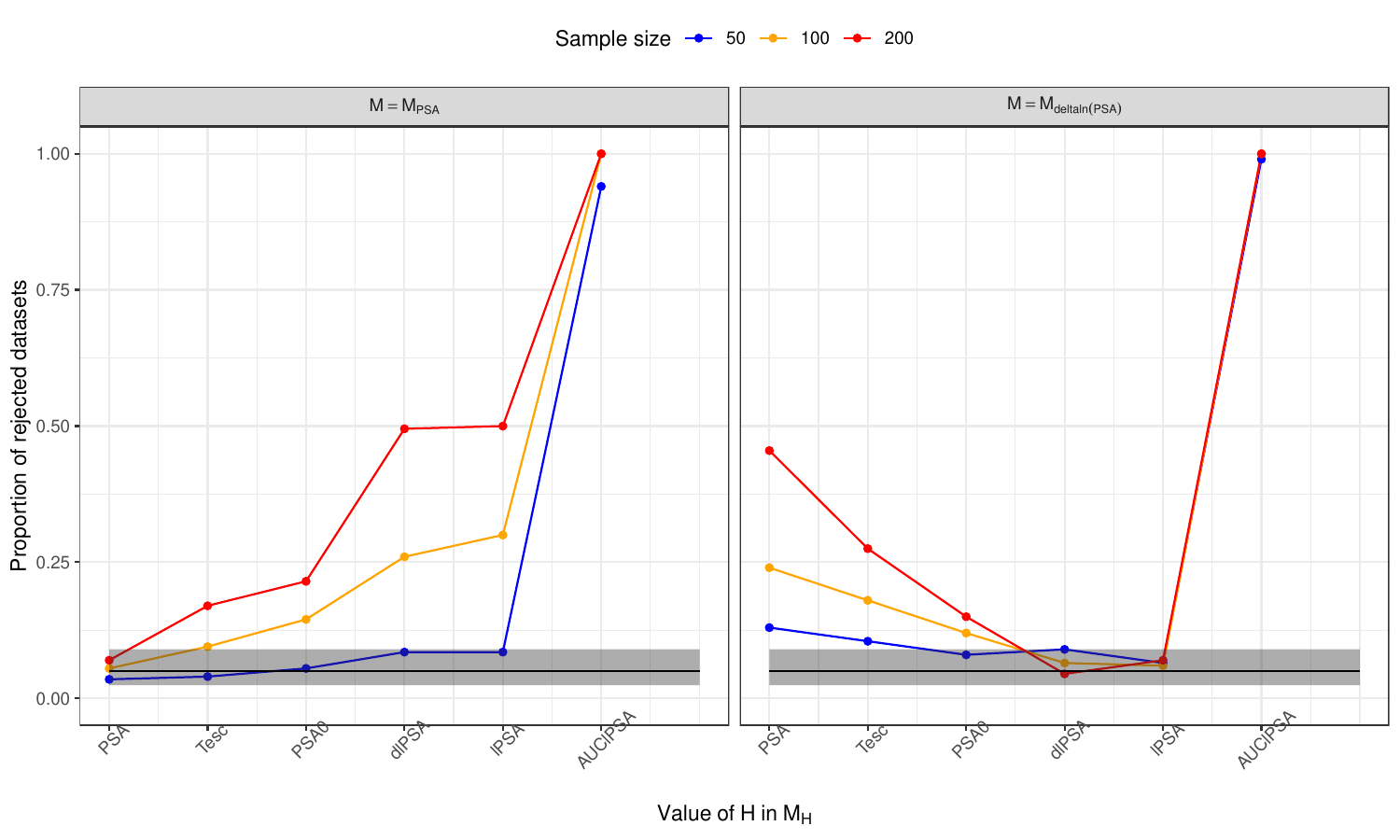}
    \caption{Performance in case of misspecifications on the association structure between both outcomes. Data are generated under a association between the current value of PSA and the hazard ($M_{PSA}$) on the left and a association between the slope of the log-PSA ($M_{\delta ln(PSA)}$) on the right. For each scenario, three sample size are tested (N=50 (blue), N=100 (orange), N=200 (red)).}
    \label{fig:power_link}
\end{figure}

\hskip 18pt Fig. \ref{fig:power_link} shows the performance when testing for misspecification on the association between both submodels. In both cases, the type I error is controlled and the power increases as the sample size increase which is, again, expected. Depending on the tested model compared to the true one, the power to detect the misspecification can be very low, for instance in the comparisons $\{M_V=M_{PSA};~ M=M_{T_{esc}}\}$, $\{~M_V=M_{\delta ln(PSA)};~M=M_{PSA_0}\}$ or $\{~M_V=M_{\delta ln(PSA)};~M=M_{ln(PSA)}\}$. On the other hand, the power to detect a model misspecification when $M=M_{AUC_{ln(PSA)}}$ is very high in this simulation for all alternatives.
    
\begin{figure}[h!]
    \centering
    \includegraphics[width=0.9\textwidth]{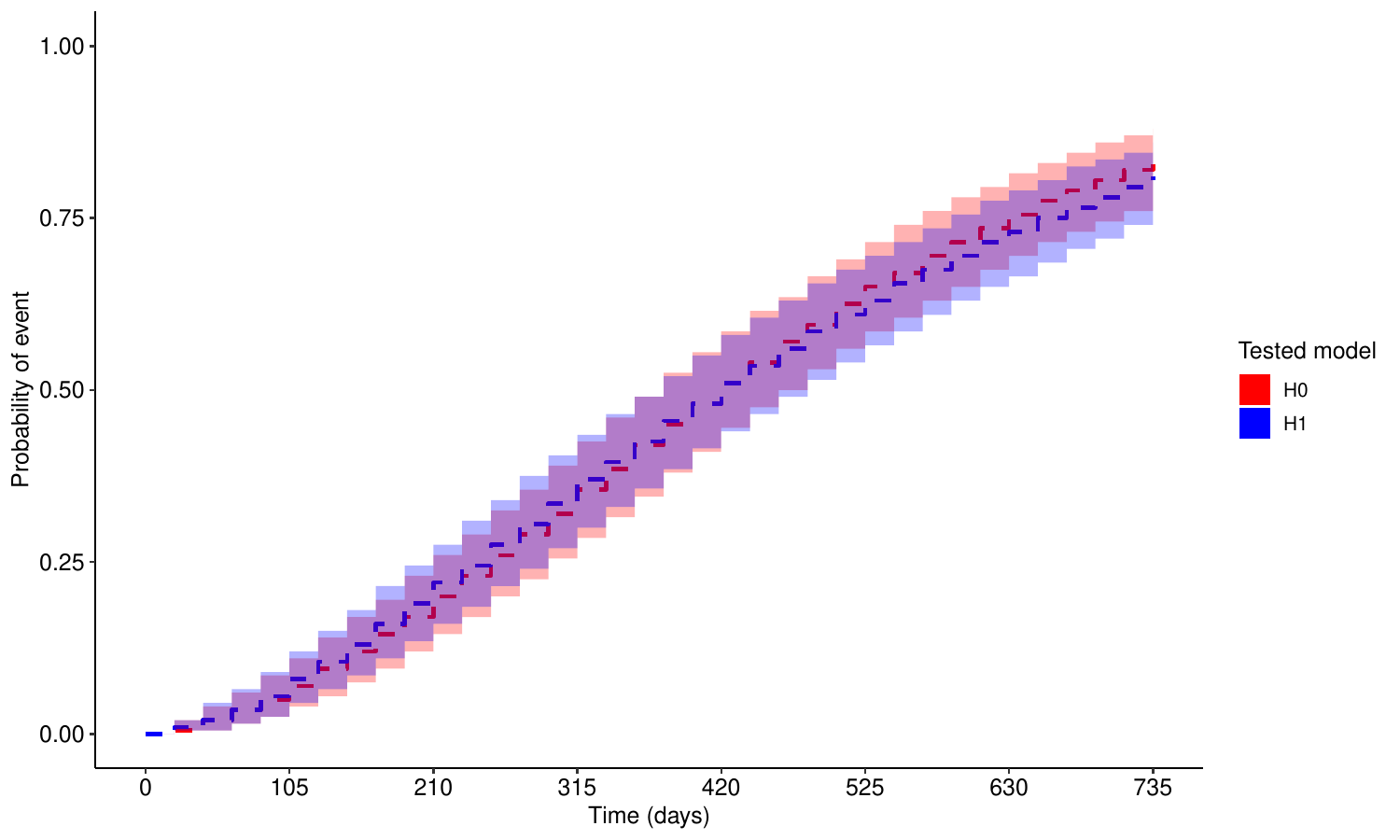}
    \caption{Kaplan-Meier VPC for the model where the association depends on the current value of PSA ($M=M_{PSA}$, in red) and the model where the association depends on the time to treatment escape ($M=M_{T_{esc}}$, in blue). The predicted median is represented with its 90\% prediction interval. }
    \label{fig:VPCH0H1}
\end{figure}

Fig. \ref{fig:VPCH0H1} provides some insight into the low power observed in some combinations, by showing the median KMVPC (with its 90\% prediction interval) of a model where the association depends on the current value of PSA ($M=M_{PSA}$, in red) and a model where the association depends on the time to treatment escape ($M=M_{T_{esc}}$, in blue), one of the cases where the power to detect a difference was very low with $npd$. The prediction intervals of the median event probability overlap over time between both models, suggesting the survival is virtually indistinguishable.





\section{Discussion}


\hskip 18pt Joint models allow the estimation of the association between processes of interest. These models can then be applied to make clinical trial simulations and support decision.  
Model evaluation of joint models with NLME and parametric TTE components is thus an important part in the model development to ensure they are able to reproduce and predict the observed data. Evaluation of models for continuous longitudinal responses has received much attention in the past decades. Evaluation methods use residual-based diagnostics, such as the weighted residuals (WRES) or the conditional weighted residuals (CWRES)~\cite{hooker_conditional_2007}, and simulation-based residuals, such as Visual Predictive Checks (VPC)~\cite{holford_visual_2005}
and $npd/npde$~\cite{mentre_prediction_2006, brendel_metrics_2006}. Simulation-based diagnostics are considered to be the gold standard~\cite{nguyen_model_2017}. We previously extended the $npd$ for TTE data with parametric models by considering the general case of interval censored events~\cite{cerou_development_2018}, using an imputation approach suggested for $npde$ in the presence of BLQ data~\cite{nguyen_extension_2012}. We showed the good performance of two distribution tests $\mathcal{N}(0,1)$ (adjusted Kolmogorov-Smirnov and global test) in the right censoring context. 


In the present paper, we show how to combine $npd_L/npde_L$ and $npd_{TTE}$ to evaluate joint models using separate tests for the two submodels, and a Bonferroni correction to combine them. We recommend the use of the $npde_L/npd_{TTE}$ for the computation of the test and the use of the $npd_L/pd_{TTE}$ for graphical evaluation. We evaluated the performance of two global adjusted tests based either on the global test or the Kolmogorov-Smirnov test.

We evaluated the performance of these tests in several misspecification settings, in the model parameters (both in the longitudinal and in the TTE part) as well as in the association function between the longitudinal marker and the event. In all cases, we found a controlled type I error and a profile of power which increases as expected as a function of the sample size and the discrepancy between the model and the data. 

The two adjusted tests had similar performances, except in the scenario where we simulated a misspecification on the between subject variability for one parameter. In this case, we found a significant loss of power with the adjusted Kolmogorov-Smirnov test. In our different simulations, rejection of the adjusted Kolmogorov-Smirnov test was strongly associated with a rejection of the global adjusted test due to the median test (Wilcoxon) on the longitudinal submodel (results not shown). On the other hand, the adjusted Kolmogorov-Smirnov test appeared less powerful when the structural process was well specified and the misspecification concerned only the variability. 

Another scenario where we observed very low power, this time regardless of the global test used, was in some cases of a misspecified association function between the longitudinal marker and the event.
Deriving prediction intervals of Kaplan-Meier plots from simulated data obtained under both models showed that the survival function was very similar for both models in these scenarios, explaining this low power. As an example, the predicted survival in our simulations was virtually indistinguishable when using the current value of the biomarker or the logarithm of that value, suggesting the two TTE models were interchangeable here. One of the limitations of these diagnostics is therefore that they are necessary to assess the adequacy of the model but not sufficient. Indeed, a good adequacy based on sufficient statistics fit by the model, such as the observed data, is not necessarily equivalent to a good extrapolation capacity~\cite{yano_evaluating_2001}. For example, in the context of dynamic individual predictions, Ferrer et al.~\citep{ferrer_individual_2019} have shown that a misspecified association function leads to a significant loss in the predictive capacity of the model. Evaluating models in terms of their ability to predict long-term outcomes of interest is therefore important for practical applications.

For the evaluation of the parametric TTE submodel, the traditional representation of 
$pd$ versus individual observation times is uninformative for TTE models, because here we consider single-event model and $pd$ measures the position of an observation within the population distribution, which is confounded with time. Instead, we propose to compute the de-trended $pd_{TTE}$ with a de-trended prediction interval. This representation is inspired from the wormplots proposed in the context of evaluation of generalised additive models~\cite{vanbuuren_wormplot_2001}, replacing the normal distribution of residuals with the uniform distribution of the $npd$. Graph of de-trended $pd_{TTE}$ showed good ability to evaluate TTE models under $H_0$ and $H_1$ by informing if there is an over/under prediction which can be interpreted: the difference is in the scale of the probability of event. These graphs may however be somewhat tricky to interpret, as a trend remaining within the prediction interval does not indicate misspecification. An alternative could be to express the $npd_{TTE}$ over the predicted median event time for each individual, which would take into account the longitudinal component. The issue with this alternative representation is that the range of the predicted median event time for all patients can be  both very different across the population and different from the time scale of the trial, and thus again this graph could be very difficult to interpret in case of model misspecification.

Graphs of $npd$ over time can inform about deviations to the model, and we simulated misspecifications on different components, the TTE submodel, the PSA submodel or the association function. The present study is set in the context of external evaluation where we assume we are trying to evaluate whether a model developed on a building dataset is able to describe appropriately the data in a validation dataset~\cite{cerou_development_2018}. Parameters were therefore not re-estimated, and for instance in the first scenario (Fig. \ref{fig:npde_npd_H0_H1_k}, middle) the PSA submodel remains correct. This highlights the importance of graphically evaluating the two submodels. The graphs and tests we propose here can also be applied for internal evaluation, however in this case the parameters for both submodels are jointly estimated and both parts can have an impact an all parameters. Thus, a misspecification on the longitudinal submodel could affect the descriptive ability of the TTE submodel even if the hazard model is correctly specified. The opposite is also true but with a lower impact since there is more information in the longitudinal process. On the other hand, re-estimating the parameters of the model can also mask possible model misspecifications, especially when the power to detect model misspecification is already low (as for instance in some misspecified association functions).

In the present work, we use the $npd_{TTE}$ to assess the parametric TTE model. Other residuals can be used for that purpose, such as Cox-Snell residuals~\cite{cox_general_1968}, martingale~\cite{therneau_martingale-based_1990} and deviance~\cite{farrington_residuals_2000} residuals. The specificity of these models is that it is not possible to fit the random effects as only one observation by individual is available: only the population parameters are estimated. Cox-Snell residuals~\cite{cox_general_1968}, used for overall model evaluation, correspond to the cumulative hazard up to the observed event time. When there is no random effect in the model, $pd$ correspond to one minus the survival function (see supplementary material), but for joint models with shared random effect, the computation of Cox-Snell residuals involves the individual parameters, whereas the $npd$ consider the predictive distribution for the entire model and is thus more appropriate for overall evaluation. We did not evaluate the Cox-Snell residuals in the present study because for the single TTE model that we used, there are no random effects on the TTE part so they are completely determined by the individual parameters estimated from the longitudinal marker. In addition, Cox-Snell residuals follow an exponential distribution under the null hypothesis but the distribution is truncated in case of censoring, which makes it difficult to compute a test and compare its performance with the $npde$. The same difficulty arises with the other residuals, which don't have a well defined distribution under $H_0$, and for deviance residuals may not be necessarily distribute around zero. They still can be useful to detect outliers and assessing the functional form but suffer the same limits as the Cox-Snell residuals.

For global model evaluation, we can use the same distinction between residual and simulation-based residuals as for continuous outcome models~\cite{nguyen_model_2017}. $npd$ are a form of simulation-based diagnostics for TTE model, and we can also produce various types of VPC on Kaplan-Meier curves. The overall evaluation of TTE model in the context of joint models is typically done with the the un-stratified KMVPC. The construction of this graph consists, as in VPC, in computing a statistic on the original data and comparing it to its distribution obtained by simulations. Recently, it has been recommended to inspect the descriptive ability not only on the survival scale but also on the hazard scale, through the hazard based VPC, as models are usually defined on the hazard scale~\cite{huh_application_2016}. The main limit with such diagnostics is that they can be difficult to compute when dealing with unbalanced or sparse design as the summary measure is computed on the population scale. In the other hand, $npd_{TTE}$ are computed for each individual with its own design. However, $npd_{TTE}$ like its simulation-based counterpart still require simulating data under the model, which can be challenging in some cases, for instance when the study involves adaptive dosing regimens or observational protocols. Another issue is the censoring process. As with other residuals, we assume that censoring is either uninformative or accounted for in the model, but it would be interesting to assess the impact of censoring mechanisms such as a study closing after a predetermined number of events.




\section{Conclusion}

In conclusion, we proposed an extension of $npd/npde$ for the evaluation of joint models with longitudinal data and TTE combining tests on both submodel components using an appropriate correction. They performed well in the simulation study investigating various model forms of misspecified models, with a controlled type I error and a power that increases with a larger difference in the shape of the survival function. We provided graphs on TTE data with interpretable information to evaluate the model, which can be useful for the construction of joint models.

We developed the $npde_L/npd_{TTE}$ for joint models, with outcomes conditionally independent with respect to random effects, and thus data missing not at random on the longitudinal part. The difficulty with joint models is that events censor the longitudinal outcome so that the marker is not observed after the event occurs, which requires an adjustment to compute the predictive distributions. On the other hand, combining the tests on different outcomes using a Bonferroni correction is a general approach which can be used more generally for multiple response models with shared random effects such as classical pharmacokinetics/pharmacodynamics models. Finally, an extension to repeated events would allow for more informative graphs involving interindividual variability in the risk of experiencing an event.

\section*{Acknowledgements}

{\bf Funding:} Marc Cerou received funding from Institut de Recherches Internationales Servier for this work, as part of a PhD research fellowship programme. Code implementing the npde in joint models is available on request.

\newpage
\renewcommand\refname{References}
\bibliographystyle{plainnat}
\bibliography{biblio}

\clearpage

\section*{Tables}

\begin{table}[h!]
\centering
\caption{\label{psa_paramter} Values of the population PSA parameters used for the simulation in all scenarios and chosen base parameter value for the TTE model, where no IIV was assumed for the parameters.}
\begin{tabular}{cccc}
\hline
Parameter & Fixed effects & Transformation & Inter-individual standard deviation ($\omega$) \\
\hline
$r$ & 0.05 & log-normal & 0.1\\
$PSA_0$ & 80 & log-normal & 0.6\\
$\epsilon$ & 0.3 & logit-normal & 1.5\\
$T_{esc}$ & 140 & log-normal & 0.6\\
$k$ & 1.5 & log-normal & -\\
$\lambda$ & 580 & log-normal & -\\
$\beta$ & 0.001 & log-normal & -\\
\hline
\end{tabular}
\end{table}

\begin{table}[h!]
    \centering
    \caption{Models for the association function $X$}
    \begin{tabular}{lcl}
    \hline
    Model & Denomination & Equation\\
    \hline
    current PSA & $M_{PSA}$ & $X(t,\psi_i) = PSA(t,\psi_i)$\\
    individual time-to-treatment escape &$M_{T_{esc}}$ & $X(t,\psi_i) = T_{esc,i}$ \\
    individual baseline PSA & $M_{PSA_{0}}$& $X(t,\psi_i) = PSA_{0,i}$\\
    slope of the logarithm of PSA & $M_{\delta ln(PSA)}$ & $X(t,\psi_i) = \frac{dPSA(t,\psi_i)}{dt}$\\
    current logarithm of PSA &$M_{ln(PSA)}$& $X(t,\psi_i) = log(PSA(t,\psi_i)+1)$\\
    cumulative logarithm of PSA & $M_{AUC_{ln(PSA)}}$& $X(t,\psi_i) = \int_0^t log(PSA(x,\psi_i)+1)dx$\\
    \hline
    \end{tabular}
    
    \label{tab:models}
\end{table}

\newpage
\section*{Figures}

\end{document}